\documentclass[12pt]{article}

\usepackage{amsfonts}
\usepackage{amsmath}
\usepackage{epsfig}
\usepackage{latexsym}
\usepackage{graphicx}
\numberwithin{equation}{section}
\allowdisplaybreaks

\usepackage[utf8]{inputenc}

\def\be{\begin{equation}}
\def\ee{\end{equation}}
\def\bea{\begin{eqnarray}}
\def\eea{\end{eqnarray}}
\newcommand{\G}{\Gamma}

\def\half{{1\over 2}}
\def\Tr{\mbox{Tr}}
\def\del{\partial}
\def\nn{\nonumber}

\begin{document}

\renewcommand{\thefootnote}{\fnsymbol{footnote}}
\setcounter{footnote}{1}


\hfuzz=100pt

\title{\begin{flushright}
\vspace{-3cm}
{\normalsize YITP-19-100}
\end{flushright}
\vspace{1.8cm}
{\Large \bf{Quantum Holographic Entanglement Entropy to All Orders in $1/N$ Expansion}}}
\date{}

\author{Shinji Hirano\footnote{shinji.hirano@wits.ac.za}
  \\ \\
{\small{\it School of Physics and Mandelstam Institute for Theoretical Physics}}
\\{\small{\it \& DST-NRF Centre of Excellence in Mathematical and Statistical Sciences (CoE-MaSS) }}
\\{\small{\it University of the Witwatersrand, WITS 2050, Johannesburg, South Africa}}\\ 
{\small\it \&}\\
  {\small{\it Center for Gravitational Physics,
Yukawa Institute for Theoretical Physics}}\\
  {\small{\it Kyoto University, Kyoto 606-8502, Japan }}
\\
}
\date{}

\maketitle

\centerline{}

\begin{abstract}
We study holographic entanglement entropy in four-dimensional quantum gravity with negative cosmological constant. By using the replica trick and evaluating path integrals in the minisuperspace approximation, in conjunction with the Wheeler-DeWitt equation, we compute quantum corrections to the holographic entanglement entropy for a circular entangling surface on the boundary three sphere. 
Similarly to our previous work on the sphere partition function, the path integrals are dominated by a replica version of asymptotically AdS conic geometries at saddle points. As expected from a general CFT argument, the final result is minus the free energy on the three sphere which agrees with the logarithm of the Airy partition function for  the ABJM theory that sums up all perturbative $1/N$ corrections despite the absence of supersymmetries.
The all-order holographic entanglement entropy cleanly splits into two parts, (1) the $1/N$-corrected Ryu-Takayanagi minimal surface area and (2) the bulk entanglement entropy across the minimal surface, as suggested in the earlier literature.
It is explicitly shown that the former comes from the localized conical singularity of the replica geometries and the latter from the replication of the bulk volume. 
\end{abstract}

\renewcommand{\thefootnote}{\arabic{footnote}}
\setcounter{footnote}{0}

\newpage

\section{Introduction}

There is a growing belief that the key to the understanding of quantum gravity is to decode the connection between geometry and information theory \cite{Almheiri:2014lwa, Pastawski:2015qua}. The precursor of this idea was seeded in the earlier works on eternal black holes \cite{Maldacena:2001kr} and holographic entanglement entropy \cite{Ryu:2006bv} in the framework of the AdS/CFT correspondence \cite{Maldacena:1997re} -- a concrete realization of holography \cite{tHooft:1993dmi, Susskind:1994vu}. 

In this work, we focus on one aspect of it and wish to add new data to the understanding of holographic entanglement entropy (HEE) proposed by Ryu and Takayanagi (RT) \cite{Ryu:2006bv} (which was further developed into a covariant form in  \cite{Hubeny:2007xt}. See \cite{Lewkowycz:2013nqa} for a proof of the RT conjecture, \cite{Nishioka:2009un} for an overview of HEE and  \cite{VanRaamsdonk:2016exw} for its potential roles in the connection between geometry and information theory.)
In particular, we provide an example of quantum gravity or $1/N$ corrections to holographic entanglement entropy. Despite the fact that the general structure of quantum corrections was proposed by Faulkner, Lewkowycz and Maldacena \cite{Faulkner:2013ana} (see also \cite{Swingle:2014uza, Barrella:2013wja}), to our knowledge, little has been known about quantum corrections to HEE. Even though our study is specific and rather formal, we hope that this work provides useful data for a more physical and universal understanding of quantum HEE.

We consider the AdS$_4$/CFT$_3$ correspondence as a specific example and adopt the path integral method, in conjunction with the Wheeler-DeWitt equation, to four-dimensional quantum gravity with negative cosmological constant. This is based on a similar method employed in the earlier works in the context of quantum cosmology \cite{Halliwell:1988wc, Halliwell:1988ik} and the same approach was applied to our previous work on the $S^3$ partition function \cite{Caputa:2018asc}. In order to compute quantum corrections to HEE, we use the replica trick and thereby generalize our previous work to the case of replica geometries.\footnote{A similar method was later used in the study of $TT$ deformations in various dimensions \cite{Caputa:2019pam} and the $T\bar{T}$ deformation and quantum HEE in AdS$_3$/CFT$_2$ \cite{Donnelly:2019pie}. The latter work has some overlap with the idea of this work.}
More specifically, we compute the entanglement entropy for a circular entangling surface on the boundary three sphere. The path integrals are evaluated in the minisuperspace approximation and reduce to a sum over saddle point geometries, which are a replica generalization of asymptotically AdS conic geometries, in a similar way to the case of the $S^3$ partition function. 

As expected from a general CFT argument, the final result \eqref{totalHEE} is minus the free energy on the three sphere which agrees with the logarithm of the Airy partition function for  the ABJM theory \cite{Aharony:2008ug, Aharony:2008gk} that sums up all perturbative $1/N$ corrections  \cite{Fuji:2011km, Marino:2011eh} despite the absence of supersymmetries.
The all-order holographic entanglement entropy cleanly splits into two parts, (1) the $1/N$-corrected Ryu-Takayanagi minimal surface area \eqref{qRT} and (2) the bulk entanglement entropy across the minimal surface \eqref{bulkEE}, as suggested in \cite{Faulkner:2013ana}.
It is explicitly shown that the former comes from the localized conical singularity of the replica geometries and the latter from the replication of the bulk volume.

The organization of this paper is as follows: In Section \ref{SPF} we review the path integral method, in particular, the evaluation of the path integrals in the minisuperspace approximation, used to compute the $S^3$ partition function in our previous work \cite{Caputa:2018asc}. Within this framework, we also provide an alternative and complementary approach using the Wheeler-DeWitt equation to bolster the weakness of the path integral method.
In Section \ref{Replica}, in order to set up the computation of quantum corrections to HEE by the replica trick, we generalize the $S^3$ partition function to the replica case, i.e. the partition function for the boundary $n$-tiply wound three sphere.
In Section \ref{HEE} we compute the all-order quantum HEE in the expasion of Newton's constant $G_N$, or equivalently, $1/N^2$. We then distill the two distinct contributions; the quantum-corrected Ryu-Takayanagi minimal surface area and the bulk entanglement entropy across the minimal surface.
In Section \ref{Discussion} we summarize our results and end with some discussions on the results and the future directions. 
Some of the technical details are relegated to Appendices \ref{Ddim} and \ref{appendix:replica}.

\section{Review of $S^3$ partition function from path integrals}
\label{SPF}

The computation of quantum HEE in this paper is a generalization of our previous work on the sphere partition function \cite{Caputa:2018asc}. We thus first review the path integral method in our previous work used to calculate the $S^3$ partition function. 

\subsection{Minisuperspace approximation}
\label{MSS}
In the path integral approach to 4$d$ quantum gravity, we integrate over all Euclidean metrics with a boundary condition of our interest, i.e. the $S^3$ boundary. We apply the method used in quantum cosmology \cite{Halliwell:1988wc, Halliwell:1988ik} to asymtotically AdS spaces with the round $S^3$ boundary.

In the Arnowitt-Deser-Misner (ADM) decomposition the general metric can be parametrized as \cite{Arnowitt:1959ah}
\be
ds^2=L^2dr^2+\gamma_{\mu\nu}\left(dx^\mu+S^\mu dr\right)\left(dx^\nu+S^\nu dr\right),
\ee
where $r$ is the radial coordinate and $x^\mu$ are the coordinates of the 3$d$ Euclidean space. $L(r, x^{\mu})$ and $S^{\mu}(r, x^{\mu})$ are the lapse and shift functions, respectively, and $\gamma_{\mu\nu}(r, x^{\mu})$ is the 3$d$ metric.
Unlike the standard ADM decomposition, the radial coordinate $r$ plays the role of the Euclidean time. 
Thus the Cauchy surfaces are ``timelike'' so that the ADM decomposition is adapted to the holographic study. 
We now decide for ourselves to work in the mimisuperspace approximation. Namely, we restrict the space to be spherically symmetric:
\be
ds^2=L^2(r)dr^2+a^2(r)d\Omega^2_{3},\label{mSS}
\ee
where $a(r)$ is the scale factor and $d\Omega^2_{3}$ is the metric on the round $S^3$. It should be noted that this is a bold approximation to make. It is far from obvious if and how accurately it can encapsulate quantum gravity effects. As we will see, however, the result provides a posteriori justification and strongly suggests that the minisuperspace approximation works surprisingly well in this case.   

The path integrals require a careful treatment involving gauge fixing and ghosts even in the minisuperspace approximation.  (See e.g. \cite{Halliwell:1988wc} for details.) Fortunately, it is well-known that after these steps are taken, one is left with path integrals over the laps $L(r)$ and the scale factor $a(r)$, 
\be
Z=\int DL \int Da\, e^{-S_E[L,a]} 
\ee
where 
\be
S_E[L,a]=S_{EH}+S_{GH}+S_{ct}\label{schematicaction}
\ee
is the regularized finite action, as we will elaborate on below.
The Euclidean action $S_{EH}$ with negative cosmological constant is the standard 4$d$ Einstein-Hilbert action and $S_{GH}$ is the Gibbons-Hawking-York boundary term \cite{Gibbons:1976ue, York:1972sj}. (See Appendix \ref{Ddim} for details and conventions.) For the metric \eqref{mSS}, after we integrate over the $S^3$ coordinates, the action can be written as
\bea
S_{EH}+S_{GH}=-\frac{V_3}{8\pi G_N}\int dr L\left[3a\left(1+\frac{a'^2}{L^2}\right)-\Lambda a^3\right],
\eea
where $V_3=2\pi^2$ is the area of the unit 3-sphere and the cosmological constant $\Lambda=-3/\ell^2$ with the AdS radius $\ell$.

The next important step is to transform the \lq\lq kinetic term'' of the scale factor into the canonical form. This is a well-known step in the study of the Hartle-Hawking wave function for the de Sitter space \cite{Hartle:1983ai, Halliwell:1988ik}, and it proves to be crucial in our discussion too. This transformation is done in two steps: First, we rescale the laps function $L\to L/a$ and then introduce a new variable $q=a^2$ so that the action becomes
\be
S_{EH}+S_{GH}=-\frac{3V_3}{8\pi G_N\ell^2}\int dr \left[\frac{\ell^2q'^2}{4L}+L\left(q+\ell^2\right)\right].\label{Actionq}
\ee
This is our main object in the following analysis and the gravitational path integrals are over the laps $L$ and $q$. In terms of the new variable, the metric takes the form
\begin{align}
ds^2&={L(r)^2 \over q(r)}dr^2 + q(r)d\Omega_3^2\ .
\end{align}
Meanwhile, the last term $S_{ct}$ in \eqref{schematicaction} is the boundary counter-term action to cancel divergences coming from the near-boundary contributions \cite{Balasubramanian:1999re, Emparan:1999pm, deHaro:2000vlm}. It is a local action composed of the induced boundary metric and given by \eqref{CTa} in Appendix \ref{Ddim}. In the minisuperspace approximation it reads
\be
S_{ct}=\frac{V_3}{8\pi G_N \ell}\left(2q_{\infty}^{3/2}+ 3\ell^2q_{\infty}^{1/2}\right) \label{SCT}
\ee
at the cutoff boundary $q=q_{\infty}\gg \ell^2$. This completes a brief account of the minisuperspace approximation.

\subsection{The $S^3$ partition function}
\label{PT}

As shown in our previous work \cite{Caputa:2018asc}, remarkably, the 4$d$ pure Einstein gravity in the minisuperspace approximation suffices to reproduce the Airy function  in the $S^3$ partition function of the ABJM theory at strong coupling \cite{Fuji:2011km, Marino:2011eh}.\footnote{There has been great progress in the understanding of the $S^3$ partition function and Wilson loops in the ABJM theory thanks to the localization techinique \cite{Drukker:2010nc, Marino:2011nm}.
The Airy partition function, including higher derivative corrections, has also been reproduced by numerical studies \cite{Honda:2012bx, Putrov:2012zi}. Moreover, non-perturbative corrections in the $1/N$ expansion have been well understood to a remarkable degree \cite{Drukker:2011zy, Hatsuda:2012dt, Hatsuda:2013oxa}. In the case of $k=1, 2$ when SUSY enhances to ${\cal N}=8$, the fully nonperturbative exact partition function has been found in \cite{Codesido:2014oua}.}

From the recap in Section \ref{MSS}, the sphere partition function in the minisuperspace approximation is given by
\begin{align}
Z_{\rm G}(S^3)=\int {\cal D}L\int {\cal D}q\exp\left[\frac{3V_3}{8\pi G_N\ell^2}\int dr \left(\frac{\ell^2q'^2}{4L}+L\left(q+\ell^2\right)\right)-S_{ct}\right]\ .
\end{align}
As an important remark, we have not specified what exactly we mean by the integration measure ${\cal D}q$. This point will be clarified along the way. Now, we choose the gauge in which the lapse $L$ is a constant \cite{Halliwell:1988wc, Halliwell:1988ik}. Our strategy is to first perform the $q$-integral in the saddle point approximation. Since the Lagrangian does not explicitly depend on \lq\lq time'' $r$, the saddle point equation is given by the \lq\lq energy'' conservation:
\begin{align}
E={\ell^2\over 4L^2} q'^2-q-\ell^2\ .\label{saddleEqn}
\end{align}
By setting $E=q_0-\ell^2$, the saddle point equation is solved to
\begin{align}
q=\left({L(r-r_0)\over\ell}\right)^2-q_0\ ,
\label{eqn:saddle}
\end{align}
where $r_0$ is an unphysical constant corresponding to the origin of \lq\lq time'', which can be shifted away, whereas $q_0$ is a parameter that characterizes each saddle point. We thus have a series of saddle point geometries labeled by $q_0$. 
\begin{figure}[h!]
\centering \includegraphics[height=1.1in]{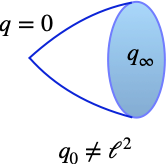} 
\hspace{3.5cm}
\includegraphics[height=1.1in]{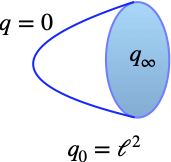} 
\caption{The saddle point geometries: The left represents generic off-shell conic geometries with $q_0\ne \ell^2$ that do not satisfy the Hamiltonian constraint. There is a curvature singularity at the tip of the cone $q=0$. The right corresponds to the special point $q_0=\ell^2$ that is on-shell and Euclidean AdS$_4$. } 
\label{saddle}
\end{figure}
They are asymptotically AdS$_4$ with the round $S^3$ boundary. Except for $q_0=\ell^2$, they are generically conic and off-shell in the sense that they do not satisfy the Hamiltonian constraint, and there is a curvature singularity at the tip of the cone $q=0$ as shown in Figure \ref{saddle}.
At any rate, the path integrals of the sphere partition function are essentially reduced to the sum over all the saddle point geometries, i.e. the integral over $q_0$.\footnote{The metric near $q=0$ takes the form $ds^2\simeq 2L\ell/\sqrt{q_0}\left(du^2+a u^2d\Omega_3^2\right)$ with $a=2L\ell^{-1}\sqrt{q_0}$. For $q_0\ne \ell^2$, the constant $a$ differs from $1$ and there is thus a curvature singularity $R\sim 1/u^2$ at $q=0$. This singularity, however, is harmless and admissible in the sense that $\sqrt{g}R\sim u$ and the minisuperspace action is finite. Moreover, as we will see, $q_0=\ell^2$ is the saddle point of the $q_0$-integral and we will argue that an appropriate choice of  the integration contour is to deform it away from the real axis at $q_0=\ell^2$. Thus the curvature singularity is avoided altogether along the complexified $q_0$-contour.}\label{footnote:singularity}

\begin{figure}[h!]
\centering \includegraphics[height=1.8in]{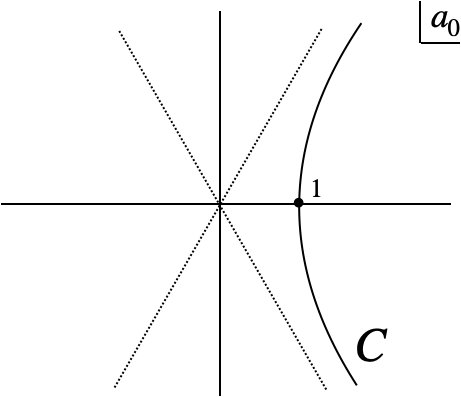} 
\caption{The contour ${\cal C}$ of the $a_0$ integration: It goes to infinity at angles $\pm \pi/3$ and passes through the saddle point $a_0=1$ (of the saddle point geometries) on the real axis. This selects the Airy function ${\rm Ai}(z)$. The parameter $q_0=\ell^2a_0^2$ of the off-shell saddle point geometries is complexified along ${\cal C}$.} 
\label{contour}
\end{figure}

It is now straightforward to find the saddle point action 
\begin{align}
\hspace{-.3cm}
S_{0}={3V_3 \over 8\pi G_N\ell}\int_{0}^{q_{\infty}} \!\!dq \left[{q_0-\ell^2\over 2\sqrt{q+q_0}}-\sqrt{q+q_0}\right]
=-{3V_3 \over 8\pi G_N\ell}\left[\frac{2}{3}q_{\infty}^{3\over 2}+\ell^2 q_{\infty}^{\half}+{1\over 3}q_0^{3\over 2}-\ell^2q_0^{\half}\right].
\label{saddleaction}
\end{align} 
The cutoff $q_{\infty}$-dependent divergences are precisely cancelled by the counter-term action $S_{ct}$ in \eqref{SCT}.
The sphere partition function, within our approximations, thus yields
\begin{align}
Z_{\rm G}(S^3)\simeq \int {\cal D}Q\int [dq_0]\exp\left[{3V_3 \over 8\pi G_N\ell}\left({1\over 3}q_0^{3\over 2}-\ell^2q_0^{\half}\right)
+\frac{3V_3}{32\pi G_N}\int dr {Q'(r)^2\over L}\right]\ ,
\label{S3partitionFull}
\end{align}
where $Q(r)$ is the fluctuation of $q(r)$ about the saddle point. By introducing a new variable $a_0$ via
\be
q_0=\ell^2 a_0^2\ ,\label{reparametrization}
\ee
we shall choose the integration measure to be $[dq_0]\propto da_0$. This will be justified a posteriori by requiring the consistency with the Wheeler-DeWitt equation, as will be discussed in Section \ref{WDW}.
As an upshot, with this choice of the measure, we obtain
\begin{align}
Z_{\rm G}(S^3)\propto {1\over 2\pi i}\int_{\cal C} da_0\exp\left[{3V_3\ell^2\over 8\pi G_N}\left({1\over 3}a_0^{3}-a_0\right)\right]
\propto {\rm Ai}\left[\left({3V_3\ell^2\over 8\pi G_N}\right)^{2\over 3}\right]\ .
\label{S3partition}
\end{align}
We chose a contour ${\cal C}$ for which the exponentially growing component of the Airy function ${\rm Bi}(z)$ is absent as shown in Figure \ref{contour}. However, for example, in the de Sitter case, both the  Hartle-Hawking \cite{Hartle:1983ai} and Vilenkin \cite{Vilenkin:1986cy} wavefunctions have the ${\rm Bi}(z)$ component.
In our case, the AdS/CFT lends strong support for this particular choice of the contour ${\cal C}$. Note that the saddle point geometries are complexified along 
${\cal C}$.

As the main result of \cite{Caputa:2018asc}, it should be stressed that the minisuperspace partition function \eqref{S3partition} is precisely the Airy function \cite{Fuji:2011km, Marino:2011eh} that appears in the $S^3$ partition function of the ABJM theory at large $\lambda$ via the AdS/CFT dictionary, $3V_3\ell^2/(8\pi G_N)=\pi N^2/\sqrt{2\lambda}$, where $N$ is the rank of the gauge groups and $\lambda=N/k$ is the 't Hooft coupling of the ${\cal N}=6$ $U(N)_k\times U(N)_{-k}$ Chern-Simons-matter theory.

A few further remarks are in order: 
(1) The saddle point parameter $a_0$ is identified with the chemical potential $\mu$ of the grand partition function of the ABJM theory \cite{Marino:2011eh}. 
(2) With the prescription $Q\to iQ$, the integral over $Q(r)$, upon regularization, yields an irrelevant constant factor.
(3) In the case of the ABJM theory, the rank $N$ is shifted to $N-k/24-1/(3k)$ in the partition function  \cite{Fuji:2011km, Marino:2011eh}. This cannot be accounted for in our approximations, since they are due to higher curvature corrections \cite{Bergman:2009zh, Aharony:2009fc}.

This result may suggest that supergravity path integrals localize to the minisuperspace in highly symmetric cases. It is, however, not clear how this can be related to the supergravity localization in \cite{Dabholkar:2014wpa}.


\subsection{Wheeler-DeWitt equation}
\label{WDW}

It is expected that the partition function is a solution to the Wheeler-DeWitt (WDW) equation. This can also be thought of as an alternative and simpler way to find the partition function. However, it is not as obvious as it may seem how exactly the \lq\lq wavefunction of the universe'' is identified with the $S^3$ partition function.

The Hamiltonian constraint for the minisuperspace action \eqref{Actionq} yields the WDW equation  \eqref{WDWdDim} as shown in Appendix \ref{Ddim}:
\begin{align}\label{WDWeq}
\left[{d^2\over dq^2}-{9\pi^2\over 16G_N^2\ell^2}\left(q+\ell^2\right) \right]\Psi_{\rm WDW}(q)=0
\end{align}
in the unit $\hbar=1$. This is the Airy equation and solved to 
\begin{align}
\Psi_{\rm WDW}(q)=C_1{\rm Ai}\left[\left({3\pi\ell^2\over 4G_N}\right)^{2\over 3}\left(\ell^{-2}q+1\right)\right]+C_2{\rm Bi}\left[\left({3\pi\ell^2\over 4G_N}\right)^{2\over 3}\left(\ell^{-2}q+1\right)\right]\ .
\end{align}
Thus we see that by choosing $C_2=0$, the $S^3$ partition function is related to the wavefunction by  
\be
Z_G(S^3)\propto \Psi_{\rm WDW}(0)\ .
\ee
This provides a concrete realization of the idea that the CFT partition function is a solution to the WDW equation as suggested in \cite{deBoer:1999tgo, McGough:2016lol}.

\begin{figure}[h!]
\centering \includegraphics[height=0.9in]{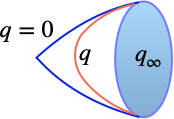} 
\caption{The IR cutoff saddle point geometries: The space is cut off at a finite $q$ (indicated by a red line) corresponding to an IR cutoff in the dual CFT, in addition to the UV cutoff at $q=q_{\infty}$.} 
\label{cutoffsaddle}
\end{figure}

In order to understand the relation between $\Psi(q)$ and the $S^3$ partition function for a generic $q$, we go back to the saddle point action \eqref{saddleaction} and cut off the space at a finite $q$, instead of going all the way down to $q=0$, as shown in Figure \ref{cutoffsaddle}:
\begin{align}
S_0\quad\longrightarrow\quad S_{0}(q)&=-{3V_3 \over 8\pi G_N\ell}\int_{q}^{q_{\infty}} dq' \left[\sqrt{q'+q_0}- {q_0-\ell^2\over 2\sqrt{q'+q_0}}\right]\nn\\
&=-{3V_3 \over 8\pi G_N\ell}\left[\frac{2}{3}q_{\infty}^{3\over 2}+\ell^2 q_{\infty}^{\half}+{1\over 3}Q_0^{3\over 2}-(q+\ell^2)Q_0^{\half}\right]
\ ,
\end{align} 
where we introduced the shifted parameter $Q_0=q_0+q$. 
Integrating over $Q_0$, the cutoff $S^3$ partition function is found to be
\begin{align}
Z_{\rm G}(S^3;q)\propto {\rm Ai}\left[\left({3V_3\ell^2\over 8\pi G_N}\right)^{2\over 3}\left(\ell^{-2}q+1\right)\right]
\propto \Psi_{\rm WDW}(q)\ .
\label{cutoffS3partition}
\end{align}
Since the radial scale $q$ corresponds to the energy scale of the dual CFT, we can interpret $q$ as an IR cutoff in the CFT and the ``wavefunction of the universe'' $\Psi_{\rm WDW}(q)$ as the IR-cutoff $S^3$ partition function, in which only the modes above the energy scale $q$ are integrated out.

As alluded in the previous sections, the choice of the $q_0$-integration measure, $[dq_0]\propto da_0$ stated below \eqref{reparametrization}, is consistent with the WDW equation and this serves as a justification for our choice of the integration measure.

\section{Replica generalization}
\label{Replica}

To apply the path integral method to the computation of holographic entanglement entropy, we are going to make use of the replica trick. Namely, we generalize the reviewed analysis of the $S^3$ partition function to the case of the multi-covered $S^3$ boundary, which we shall denote by $S^3_n$. 

\begin{figure}[h!]
\centering \includegraphics[height=1.3in]{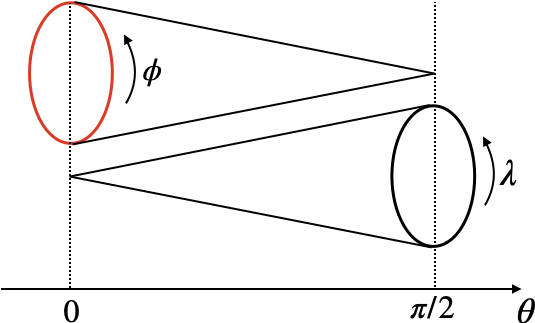} 
\caption{The (multi-covered) round $S^3_n$ and the entangling surface: The circular entangling surface is the $\phi$-circle located at $\theta=0$ indicated in red. The $\lambda$-circle winds around $n$ times, $\lambda\sim \lambda+2n\pi$, for the  $n$-covered $S^3_n$, and there is a conical singularity at $\theta=0$ where the entangling surface is located.} 
\label{replicasphere}
\end{figure}

We parametrize the round $S^3$ by 
\begin{align}
d\Omega^2_{3}=\cos^2\theta d\phi^2+d\theta^2+\sin^2\theta d\lambda^2\quad\mbox{with}\quad \theta\in [0, \pi/2]\ ,\,\,\phi\sim\phi+2\pi\ ,\,\,\lambda\sim\lambda+2\pi\ .
\end{align}
The replica $S^3_n$ of the boundary $S^3$ is obtained by winding the $\lambda$-circle $n$ times round, i.e. $\lambda\sim\lambda+2n\pi$. This creates a conical singularity at $\theta=0$ and suitable for the circular entangling surface along the $\phi$-circle located at $\theta=0$, as shown in Figure \ref{replicasphere}.

The bulk replica geometries are described by the metric
\be
ds_n^2={L^2dr^2\over q(r)}+q(r)d\Omega_{3,n}^2\ ,\label{replicametric}
\ee
where $d\Omega_{3,n}^2$ is the metric on the boundary replica sphere $S^3_n$ with $\lambda\sim\lambda+2n\pi$ and a unit radius. Note that the scalar curvature of the replica $S^3_n$ has a $\delta$-function singularity at $\theta=0$ and is given by 
\be
R_3=6+{2\over \sin\theta\cos\theta}\left({1\over n}-1\right)\delta(\theta)\ ,
\ee
which was computed, for example, in \cite{Solodukhin:1994yz} and in a more similar context \cite{Nishioka:2013haa, Nishioka:2014mwa}.
Now, using the results in Appendix \ref{appendix:replica}, we find the replica minisuperspace action 
\begin{align}\label{replicaaction}
S_{EH}+S_{GH}=-{3nV_3\over 8\pi G_N\ell^2}\int dr\left[{\ell^2q'^2\over 4L}+L\left(q+{\ell^2\over 3}\left(1+{2\over n}\right)\right)\right]\ .
\end{align}
This generalizes the $S^3$ case \eqref{Actionq} to the replica $S^3_n$. The saddle point equation \eqref{saddleEqn} is unchanged and thus the saddle point geometries are given by the same $q(r)$ as that in the $S^3$ case \eqref{eqn:saddle}.
Comparing the $S^3$ case \eqref{Actionq} and the replica case \eqref{replicaaction}, the net effects are (1) the replication of the bulk volume by a factor of $n$ and (2) a shift of the variable $\ell^{-2}q$ by
\be\label{netshift}
\ell^{-2}q+1\quad\rightarrow\quad \ell^{-2}q+1+{2\over 3}\left({1\over n}-1\right)\ .
\ee
This implies that the cutoff $S^3$ partition function \eqref{cutoffS3partition} is generalized to
\begin{align}
Z_{\rm G}(S^3_n;q)\propto {\rm Ai}\left[\left({3nV_3\ell^2\over 8\pi G_N}\right)^{2\over 3}\left(\ell^{-2}q+1+{2\over 3}\left({1\over n}-1\right)\right)\right]\ .
\label{cutoffreplicaS3partition}
\end{align}
Note that this is also the ``wavefunction of the universe'' $\Psi_{{\rm WDW}_n}(q)$ for the replica $S^3_n$ boundary, as can be seen from the straighforward generalization of the discussion in Section \ref{WDW}. Hence, we find the replica $S^3_n$ partition function to be
\begin{align}
Z_{\rm G}(S^3_n)\propto {\rm Ai}\left[\left({3nV_3\ell^2\over 8\pi G_N}\right)^{2\over 3}\left(1+{2\over 3}\left({1\over n}-1\right)\right)\right]\ .
\label{replicaS3partition}
\end{align}
It should be stressed that the effect (1) is a {\it bulk} effect and (2) is a {\it localized} effect due to the conical singularity of the replica geometries. We will elaborate on this point in the next section.
\section{Quantum holographic entanglement entropy}
\label{HEE}

We are now in a position to compute quantum corrections to holographic entanglement entropy for the circular entangling surface shown in Figure \ref{replicasphere}.
For this special entangling surface, the entanglement entropy (EE) can be found, via the replica trick, by the formula
\begin{align}\label{EEfromreplica}
{\cal S}_{EE}&=-{\del\over\del n}\log{{\cal Z}_G(S^3_n)\over {\cal Z}_G(S^3)^n}\biggr|_{n=1}\ ,
\end{align}
where the round sphere $S^3=S^3_{1}$. Here we introduced the normalized partition function
\begin{align}
{\cal Z}_G(S^3_n)={\rm Ai}\left[\left({3nV_3\ell^2\over 8\pi G_N}\right)^{2\over 3}\left(1+{2\over 3}\left({1\over n}-1\right)\right)\right]\ .
\end{align}
Strictly speaking, the minisuperspace path integrals \eqref{S3partitionFull} do not categorically explain why there does not arise a normalization factor dependent on $\ell^2/G_N$. This is a weakness of our approach. However, in conjunction with the WDW equation \eqref{WDWeq} and its straightforward generalization to the replica geometry, it is consistent and judicious to conclude that such a factor is absent.

The replica formula \eqref{EEfromreplica} then yields
\begin{align}\label{totalHEE}
{\cal S}_{EE}&=-\left(3\pi \ell^2\over 4G_N\right)^{2\over 3}\left(\underbrace{2\over 3}_{bulk}-\underbrace{{2\over 3}}_{local}\right){{\cal Z}'_G(S^3)\over {\cal Z}_G(S^3)}+\underbrace{\log {\cal Z}_G(S^3)}_{-F(S^3)\,\,\&\,\, bulk}=-F(S^3)\ .
\end{align}
As expected from a general CFT argument \cite{Casini:2010kt}, the entanglement entropy is minus the free energy $F(S^3)$ on the three sphere. 
It is worth emphasizing that this is a result to all orders in the $1/N$ expansion: In the ABJM parametrization, it is expressed as
\begin{align}
{\cal S}_{EE}=\log  {\rm Ai}\left[\left({\pi N^2\over \sqrt{2\lambda}}\right)^{2\over 3}\right]\ .
\end{align}
As remarked above and suggested in the earlier literature \cite{Faulkner:2013ana}, this consists of two kinds of  contributions: One is the {\it bulk} entanglement entropy across the Ryu-Takayanagi (RT) minimal surface and the other is the quantum-corrected RT minimal surface area {\it localized} at $\theta = 0$. The first $2/3$ term in \eqref{totalHEE} is due to the factor $n^{2/3}$ in the bulk volume factor in \eqref{replicaS3partition}, whereas the second $2/3$ term is due to $2/3\cdot 1/n$ in \eqref{replicaS3partition} which stems from the localized singularity at $\theta=0$.

We are first going to discuss the latter and then turn to the former. In addition, we will reinstate the $q_{\infty}$-dependent terms and show that the bare divergent contribution to the entanglement entropy is precisely that of the RT minimal surface.

\subsection{Quantum Ryu-Takayanagi minimal surface area}
\label{HEERT}

As indicated in \eqref{totalHEE} and explained above, the part of the entanglement entropy localized at $\theta=0$ is expected to be the quantum-corrected RT minimal surface area (divided by $4G_N$): 
\begin{align}\label{qRT}
{\cal S}_{qRT}={{\cal A}_{qRT}\over 4G_N}&={2\over 3}\left(3\pi \ell^2\over 4G_N\right)^{2\over 3}{{\cal Z}'_G(S^3)\over {\cal Z}_G(S^3)}\ .
\end{align}
We are now going to show that this is indeed the case. In order to do so, we shall first find the minimal surface and then compute the {\it quantum average} of the minimal surface area. Technically speaking, the minimal surface coincides with the string worldsheet dual to a $1/2$-BPS Wilson loop studied in our previous work \cite{Caputa:2018asc}.\footnote{Adding a string probe \cite{Maldacena:1998im, Rey:1998ik}  dual to a $1/2$-BPS circular Wilson loop \cite{Drukker:2008zx, Chen:2008bp, Rey:2008bh} to the path integral method reviewed in Section \ref{SPF}, we have precisely reproduced the Airy function part of the $1/2$-BPS Wilson loop computed by a Fermi gas approach to the ABJM theory \cite{Klemm:2012ii}.} 

\begin{figure}[h!]
\centering \includegraphics[height=0.9in]{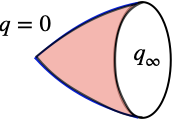} 
\caption{The RT minimal surface for the entangling $\phi$-circle at $\theta=0$: It is topologically a disk with a conic singularity at $q=0$.} 
\label{fig:RT}
\end{figure}

Since the boundary of the minimal surface is the $\phi$-circle at $\theta=0$, the induced metric on the minimal surface is given by\footnote{The fact that this is minimal can easily be understood as follows: For a generic embedding $r=r(\tau, \sigma)$ and $\phi=\phi(\tau, \sigma)$ with the worldsheet coordinates $(\tau, \sigma)$, the Nambu-Goto Lagrangian reads ${\cal L}_{NG}=L(\dot{r}\phi' - \dot{\phi} r')$. This is minimal for any embedding, and we can, in particular, choose $(r, \phi)=(\tau, \sigma)$ which gives the induced metric on the minimal surface \eqref{minsurfmetric}.}
\begin{align}\label{minsurfmetric}
ds^2_{2d}={L^2\over q(r)}dr^2+q(r)d\phi^2\ .
\end{align}
This is the 2$d$ subspace at $\theta=0$ of the 4$d$ space \eqref{replicametric} with $n=1$, i.e. the $S^3$ boundary as illustrated in Figure \ref{fig:RT}.
The minimal surface area in the saddle point geometries \eqref{eqn:saddle} is then given by
\begin{align}
{\cal A}^{(bare)}_{RT}(a_0)=\int \!dr\!\int_0^{2\pi} d\phi\sqrt{\det g_{2d}}=\pi\ell\int_0^{q_{\infty}}\!{dq\over \sqrt{q+q_0}}=2\pi\ell\left(\sqrt{q_{\infty}+(\ell a_0)^2}-\ell a_0\right)\ ,
\end{align}
where we used $q_0=\ell^2a_0^2$ as defined in \eqref{reparametrization}. The superscript ``bare'' means that the minimal area is not renormalized.
This is the {\it classical} RT minimal surface area.

We now propose that the quantum-corrected RT minimal surface is given by the {\it quantum gravity average} 
\begin{align}\label{QuantumRT}
{\cal A}^{(bare)}_{qRT}&=\left\langle {\cal A}^{(bare)}_{RT}(a_0)\right\rangle_{QG} \equiv {\cal Z}_G(S^3)^{-1}
{{\cal N}\over 2\pi i}\int_C da_0 {\cal A}^{(bare)}_{RT}(a_0)e^{{3V_3\ell^2\over 8\pi G_N}\left({1\over 3}a_0^3-a_0\right)}\nn\\
&\stackrel{q_{\infty}\to\infty}{\longrightarrow} 2\pi\ell\sqrt{q_{\infty}}- {2\pi\ell^2\over {\cal Z}_G(S^3)}
{{\cal N}\over 2\pi i}\int_C da_0\,a_0\, e^{{3\pi\ell^2\over 4 G_N}\left({1\over 3}a_0^3-a_0\right)}\ ,
\end{align}
where the normalization factor ${\cal N}=({\scriptstyle{3V_3\ell^2\over 8\pi G_N}})^{{\scriptscriptstyle{1\over 3}}}$ such that $\langle 1\rangle_{QG} = 1$.
Note that this is a different observable from Wilson loops. Formally speaking, it appears to be identical to a vev of the logarithm of Wilson loops.
As we expected, this quantum-corrected area yields the localized part of the entanglement entropy
\begin{align}\label{bareqRTentropy}
{\cal S}^{(bare)}_{qRT}\equiv {{\cal A}^{(bare)}_{qRT}\over 4G_N}= {\pi\ell\over 2 G_N}\sqrt{q_{\infty}}+{2\over 3}\left({3\pi\ell^2\over 4 G_N}\right)^{2\over 3}{{\cal Z}'_G(S^3)\over {\cal Z}_G(S^3)}={\cal S}_{div}+{\cal S}_{qRT}\ ,
\end{align}
where we defined the divergent $q_{\infty}$-dependent part of the entanglement entropy ${\cal S}_{div}= {\pi\ell\over 2 G_N}\sqrt{q_{\infty}}$. 
As advertized in \eqref{qRT}, we have thus shown that the localized part of the entanglement entropy is indeed the quantum-corrected RT minimal surface area (divided by $4G_N$).
Note that the renormalized finite part of the quantum RT entropy can be rewritten and expanded as
\begin{align}\label{neatSqRT}
{\cal S}_{qRT}=\half\ell{\del\over\del\ell}\log {\cal Z}_G(S^3)=-{2\pi\ell^2\over 4G_N}-\frac{1}{6}+\frac{5}{48}\left({4G_N\over 3\pi\ell^2}\right)+{\cal O}(G_N^2)\ ,
\end{align}
where $-2\pi\ell^2$ in the leading ${\cal O}(G_N^{-1})$ term is the regularized area of the 2$d$ hyperbolic space $H_2$, i.e. the classical RT minimal surface.

We will come back to the divergent part $S_{div}$ in Section \ref{HEEdivergences} and show that it is precisely the divergent contribution obtained via the replica trick for the unrenormalized entanglement entropy.

\subsection{Bulk entanglement}
\label{HEEbulk}

Having identified the localized part with the quantum RT minimal surface area, as indicated in \eqref{totalHEE}, the rest must be the bulk entanglement entropy across the minimal surface \eqref{minsurfmetric} as suggested in  \cite{Faulkner:2013ana}: 
\begin{align}\label{bulkEE}
{\cal S}_{bulk}&=-{2\over 3}\left(3\pi \ell^2\over 4G_N\right)^{2\over 3}{{\cal Z}'_G(S^3)\over {\cal Z}_G(S^3)}+\log {\cal Z}_G(S^3)\ .
\end{align}
Since the first and the last contributions in \eqref{totalHEE} are not localized anywhere in the bulk and come from all over the bulk space, it is very intuitive to identify them with the bulk entanglement entropy. In principle, there could be a Wald-like entropy. However, since there is no higher derivative correction present in this case, we can exclude the possibility of having a Wald-like entropy.    

Unlike the case of the localized contribution, we have no independent way to check \eqref{bulkEE} by the technology of this work. We can nonetheless do a simple check with the expected result established in the earlier literature. Namely, the leading quantum correction to the bulk contribution is expected to be logarithmic \cite{Faulkner:2013ana, Fujita:2009kw}. Indeed, we can explicitly see that
\begin{align}\label{bulkHEE}
{\cal S}_{bulk}&=\left(1-{1\over 2}\ell{\del\over \del\ell}\right)\log {\cal Z}_G(S^3)
=\frac{1}{6}\log \left(\frac{G_N}{48 \pi ^4 \ell^2}\right)+\frac{1}{6}-\frac{5G_N}{108\pi \ell^{2}}+{\cal O}(G_N^2)\ ,
\end{align}
where we used \eqref{neatSqRT}.
Note that a slightly nontrivial cancellation of the order ${\cal O}(G_N^{-1})$ terms has happened in \eqref{bulkHEE}. 

The leading logarithmic correction is identical to the universal one-loop correction in 11$d$ supergravity on $AdS_4\times X_7$ computed in \cite{Bhattacharyya:2012ye}, where $X_7$ are {\it any} compact Einstein seven-manifolds of positive curvature such as tri-Sasaki Einstein manifolds. It was shown there that the universal correction is due only to the zero mode of a 2-form ghost for the 3-form field in 11$d$ supergravity. From the pure 4$d$ gravity viewpoint, this is rather puzzling because there are only gravitons. However, as the cosmological constant can be described by a constant 4-form field strength of the 3-form field, it might be that when we reformulate the path integrals in terms of the metric plus the 3-form field, our approximations and choice of the measure can be more rigorously justified.

\subsection{Divergences}
\label{HEEdivergences}

To be complete, we derive the divergent $q_{\infty}$-dependent entropy in \eqref{bareqRTentropy} via the replica trick. Since we are now interested in the {\it bare} holographic entanglement entropy, we undo the renormalization by dropping the counter-term action $S_{ct}$. 

From \eqref{replicaaction}, it is straightforward to find the divergent part of the action for the replica geometries
\begin{align}
S_{div}(n)=-{3n\pi\over 4G_N\ell}\left[{2\over 3}q_{\infty}^{3\over 2}+{\ell^2\over 3}\left(1+{2\over n}\right)q_{\infty}^{1\over 2}\right]\ .
\end{align}
Then the replica formula computes the  $q_{\infty}$-dependent entropy as
\begin{align}
{\cal S}_{div}&=-{\del\over\del n}\log\left[e^{-S_{div}(n)+nS_{div}(1)}\right]\biggr|_{n=1}={\pi\ell\over 2G_N}\sqrt{q_{\infty}}\ .
\end{align}
This precisely agrees with the divergent part of the RT minimal surface area divided by $4G_N$ in \eqref{bareqRTentropy}.
Note that the bulk contributions from the overall factor of $n$ cancel out and what is left is the only localized contribution.
This implies that there is no divergence in the quantum corrections to the holographic entanglement entropy.

\section{Discussion}
\label{Discussion}

By using the replica trick and evaluating the path integrals in the minisuperspace approximation, in conjunction with the Wheeler-DeWitt equation, 
we computed the quantum gravity or $1/N$ corrections to holographic entanglement entropy to all orders in the $G_N$, or equivalently $1/N^2$, expansion in the AdS$_4$ gravity. Here we summarize the final results: The quantum holographic entanglement entropy is minus the free energy on the three sphere
\begin{align}
{\cal S}_{EE}=\log {\rm Ai}\left[\left({3\pi\ell^2\over 4 G_N}\right)^{2\over 3}\right]=\log  {\rm Ai}\left[\left({\pi N^2\over \sqrt{2\lambda}}\right)^{2\over 3}\right]
\end{align}
which is composed of two distinct contributions; (1) the quantum-corrected Ryu-Takayanagi minimal surface area, (Only the finite part is presented and see Section \ref{HEEdivergences} on the divergent part.)
\begin{align}
{\cal S}_{qRT}=\half\ell{\del\over\del\ell}{\cal S}_{EE}=-{2\pi\ell^2\over 4G_N}-\frac{1}{6}+\frac{5}{48}\left({4G_N\over 3\pi\ell^2}\right)+{\cal O}(G_N^2)\
\end{align}
and (2) the bulk entanglement entropy across the minimal surface,
\begin{align}
{\cal S}_{bulk}&=\left(1-{1\over 2}\ell{\del\over \del\ell}\right){\cal S}_{EE}
=\frac{1}{6}\log \left(\frac{G_N}{48 \pi ^4 \ell^2}\right)+\frac{1}{6}-\frac{5}{144}\left(\frac{4G_N}{3\pi \ell^{2}}\right)+{\cal O}(G_N^2)\ .
\end{align}
This provides a concrete example of quantum holographic entanglement entropy proposed by Faulkner, Lewkowycz and Maldacena \cite{Faulkner:2013ana}.
The minimal surface \eqref{minsurfmetric} in the saddle point geometries \eqref{eqn:saddle} coincides with the classical RT minimal surface at $q_0=\ell^2$ but is otherwise an off-shell generalization. However, it is not clear how it can be understood as a quantum extremal surface as suggested by Engelhardt and Wall \cite{Engelhardt:2014gca}. It would, of course, be very interesting to find a way to connect the minimal surface \eqref{minsurfmetric} and the quantum extremal surface of the generalized entropy.

An obvious and potentially interesting generalization of this work is to study qauntum HEE for non-maximal circular entangling surfaces, i.e. the cases of the surfaces at $\theta=\theta_0\ne 0$. In these generalizations, the boundary $S^3$ is divided into the region $A=\{(\theta, \phi, \lambda)|\, 0\le \theta \le\theta_0, 0\le\phi, \lambda\le 2\pi\}$ and its complement $\bar{A}$. Thus there may be a way to directly use the reduced density matrix $\rho_A=\Tr_{\bar{A}}\rho$, by splitting the WDW wavefunction into the $A$ and $\bar{A}$ Hilbert spaces, to calculate quantum HEE. This might shed light on and provide some insights into the meaning and interpretation of the WDW wavefunction as a ``wavefunction of the universe.''

Finally, given a rather remarkable success of our simple (and simplistic) approach to the study of quantum gravity and $1/N$ corrections, it is worthwhile to push the envelop further and test the applicability of this approach. Of particular interest is its application to the microstate counting of AdS$_4$ magnetic black holes for which the agreement between gravity and dual gauge theory results has been established at large $N$ \cite{Benini:2015eyy}. (See \cite{Zaffaroni:2019dhb} for a review and more recent developments.)


\section*{Acknowledgment}

I would like to thank Pawel Caputa for the collaboration in the early stage of this work and Masaki Shigemori for discussions. 
I would also like to thank the Graduate School of Mathematics and the Department of Physics at Nagoya University for their kind hospitality. 
This work was supported in part by the National Research Foundation of South Africa and DST-NRF Centre of Excellence in Mathematical and Statistical Sciences (CoE-MaSS).
Opinions expressed and conclusions arrived at are those of the author and are not necessarily to be attributed to the NRF or the CoE-MaSS.

\appendix
\renewcommand{\theequation}{\Alph{section}.\arabic{equation}}

\section{Computational details for minisuperspace action}\label{Ddim}
For completeness, we present a detailed derivation of the gravity action in the minisuperspace approximation as well as the WDW equation for arbitrary dimensions $d+1$.

The Euclidean gravity action is defined as
\be
S_{EH}+S_{GH}=-\frac{1}{16\pi G_N}\int_{\mathcal{M}}d^{d+1}x\sqrt{g}\left(R-2\Lambda\right)+\frac{1}{8\pi G_N}\int_{\partial \mathcal{M}}d^dx\sqrt{\gamma}\Theta
\ee
with the negative the cosmological constant 
\be
\Lambda=-\frac{d(d-1)}{2\ell^2}\ .
\ee
The minisuperspace ansatz for the metric is given by
\be
ds^2=g_{\mu\nu}dx^\mu dx^\nu=L^2(r)dr^2+a^2(r)d\Omega^2_{d}\ ,
\ee
where $d\Omega^2_{d}$ is a metric on the $d$-dimensional sphere with the volume
\be
V_{d}=\int d\Omega_d=\frac{2\pi^{\frac{d+1}{2}}}{\G\left(\frac{d+1}{2}\right)}\ .
\ee
The Ricci scalar can be expressed in terms of the laps and the scale factor as 
\be
R=d(d-1)\left[\frac{1}{a^2(r)}-\frac{a'(r)^2}{a^2(r)L^2(r)}\right]+2d\left[\frac{a'(r)L'(r)}{a(r)L(r)^3}-\frac{a''(r)}{a(r)L^2(r)}\right].
\ee
Meanwhile, the extrinsic curvature is defined by
\be
\Theta^{\mu\nu}=-\frac{1}{2}\left(\nabla^\mu\hat{n}^\nu+\nabla^\nu\hat{n}^\mu\right)\ ,
\ee
and for the boundary at constant $r$ we have the normal vectors
\be
\hat{n}^\mu=L^{-1}(r)\delta^{\mu,r}\ ,\qquad\quad g_{\mu\nu}\hat{n}^\mu\hat{n}^\nu=1
\ee
so that
\be
\Theta=-g_{\mu\nu}\nabla^\mu\hat{n}^\nu=\frac{L'(r)}{L^2(r)}-\G^\mu_{\mu r}L^{-1}(r)=-\frac{d\,a'(r)}{L(r)a(r)}\ ,
\ee
where we used the nonvanishing components of the Christoffel symbols
\be
\G^r_{rr}=\frac{L'(r)}{L(r)}\ ,\qquad\quad \G^{\theta_i}_{\theta_i r}=\frac{a'(r)}{a(r)}\ .
\ee
The Einstein-Hilbert action then becomes
\bea
&&-\frac{1}{16\pi G_N}\int_{\mathcal{M}}d^{d+1}x\sqrt{g}(R-2\Lambda)=-\frac{V_d}{16\pi G_N}\int dr L(r)\left[d(d-1)a^{d-2}(r)\left(1+\frac{a'(r)^2}{L^2(r)}\right)\right]\nn\\
&&+\frac{V_d}{8\pi G_N}\int dr L(r) \Lambda a^d(r)+\frac{V_d d}{8\pi G_N}\int dr \partial_r\left(a^{d-1}(r)\frac{a'(r)}{L(r)}\right)\ .
\eea
On the other hand, the Gibbons-Hawking-York boundary term \cite{Gibbons:1976ue, York:1972sj} 
\be
\frac{1}{8\pi G_N}\int_{\partial \mathcal{M}}d^dx\sqrt{\gamma}\,\Theta=-\frac{V_d d}{8\pi G_N}\left[ \frac{a^{d-1}(r) a'(r)}{L(r)}\right]_{\rm bdy}
\ee
precisely cancels the boundary contribution from the bulk action and we have
\be
S_{EH}+S_{GH}=-\frac{V_d}{8\pi G_N}\int dr L\left[\frac{d(d-1)}{2}a^{d-2}\left(1+\frac{a'^2}{L^2}\right)-\Lambda a^d\right]\ .
\ee
Next, for the canonical kinetic term, we first redefine the laps function $L\to La^{d-4}$ and the introduce a new variable $q=a^2$ that brings us to\\
\be
S_{EH}+S_{GH}=-\frac{V_d}{8\pi G_N}\int dr \left[\frac{d(d-1)}{2}\frac{q'^2}{4L}+L\left(\frac{d(d-1)}{2}q^{d-3}-\Lambda q^{d-2}\right)\right].\label{ActionqDd}
\ee\\
For $d=3$ this reproduces the action used in the main text.

To subtract the divergences we use the standard counter-term action \cite{Balasubramanian:1999re, Emparan:1999pm, deHaro:2000vlm}\footnote{This counter-term action is valid for $d=2,3,4$ i.e. $AdS_{3,4,5}$ and for $d=2$ i.e. $AdS_3$ we only take the first term.}
\be
S_{ct}=\frac{1}{8\pi G_N}\int_{\partial M}\sqrt{\gamma}\left(\frac{d-1}{\ell}+\frac{\ell}{2(d-2)} R_{c}(r)\right)\label{CTa}
\ee
where $\sqrt{\gamma}=a^d(r)d\Omega_d$ and the Ricci scalar of the induced metric at constant $r$ is
\be
R_c(r)=\frac{d(d-1)}{a^2(r)}\ .
\ee

Finally, we derive the Wheeler-DeWitt equation from \eqref{ActionqDd}: We first define the canonical \lq\lq momentum'' conjugate to $q(r)$
\be
p\equiv \frac{\partial L}{\partial q'}=-\frac{V_d}{8\pi G_N}\frac{d(d-1)}{4L}q'\ .
\ee
By the Legendre transformation $H=q'p-{\cal L}$, we find the \lq\lq Hamiltonian'' 
\be
H=L\hat{H}=-\frac{16\pi G_N }{V_d d(d-1)}L\left[p^2-\left(\frac{d(d-1)V_d}{16\pi G_N\ell}\right)^2\left(\ell^2q^{d-3}+q^{d-2}\right)\right].
\ee
By using the differential form of the momentum, $p=\hbar\frac{d}{dq}$, we arrive at the Hamiltonian constraint, or the Wheeler-DeWitt equation, for the wavefunction
\be
\hat{H}\Psi(q)=\left[\hbar^2\frac{d^2}{dq^2}-\left(\frac{d(d-1)V_d}{16\pi G_N\ell}\right)^2\left(\ell^2q^{d-3}+q^{d-2}\right)\right]\Psi(q)=0\ .\label{WDWdDim}
\ee
In four dimensions ($d=3$), this becomes the Airy equation. It is also intriguing to note that in 5 dimensions ($d=4$) the equation can be written in the form of the Schr\"odinger equation for a simple harmonic oscillator whose solution is given in terms of Hermite polynomials.

\section{Computational details for replica geometries}\label{appendix:replica}

In this appendix, we provide computational details for the derivation of the minisuperspace action \eqref{replicaaction} for the replica geometries.
The metric of our interest is of the form
\be
ds^2=e^{2g(r)}dr^2+e^{2h(r)}ds_3^2\ ,\label{generalmetric}
\ee
where $ds_3^2=\delta_{ab}e^ae^b$ in terms of tribein 1-forms. An efficient way to compute the scalar curvature is to first compute the curvature 2-form:
\be
R^{A}_{\mbox{ }B}=d\omega^{A}_{\mbox{ }B}+\omega^{A}_{\mbox{ }C}\wedge \omega^{C}_{\mbox{ }B}
\qquad\mbox{with}\qquad
de^A+\omega^{A}_{\mbox{ }B}\wedge e^B=0\ ,
\ee
where $e^A$ is a vierbein and labeled by the indices $A=(r, a)$ with $a=1,2,3$. For the metric \eqref{generalmetric} it takes the form
\be
e^A=\left(e^{g(r)}dr, e^{h(r)} e^a\right)\ .
\ee
The spin connection $\omega^{A}_{\mbox{ }B}$ is found by solving
\begin{align}
\omega^{r}_{\mbox{ }a}\wedge e^a=0\ ,\qquad
e^h\left(h'dr\wedge e^a+de^a\right)+e^g\omega^{a}_{\mbox{ }r}\wedge dr+e^h\omega^{a}_{\mbox{ }b}\wedge e^b=0\ .
\end{align}
These equations split into
\begin{align}
\omega^{r}_{\mbox{ }a}\wedge e^a=0\ ,\qquad
dr\wedge\left(e^h h' e^a-e^g \omega^{a}_{\mbox{ }r}\right)=0\ ,\qquad
de^a+\omega^{a}_{\mbox{ }b}\wedge e^b=0\ .
\end{align}
The first two equations can be solved to
\be
\omega^{a}_{\mbox{ }r}=-\omega^{r}_{\mbox{ }a}=e^{h-g}h' e^a\ .
\ee
The straightforward computation then yields
\begin{align}
R^{r}_{\mbox{ }r}&=0\ ,\\
R^{r}_{\mbox{ }a}&=-\left(e^{h-g}h'\right)'dr\wedge e_a-e^{h-g}h'(de_a+e_b\wedge \omega^{b}_{\mbox{ }a})=-\left(e^{h-g}h'\right)'dr\wedge e_a\ ,\\
R^a_{\mbox{ }b}&=d\omega^{a}_{\mbox{ }b}+\omega^{a}_{\mbox{ }c}\wedge \omega^{c}_{\mbox{ }b}
+\omega^{a}_{\mbox{ }r}\wedge \omega^{r}_{\mbox{ }b}
=d\omega^{a}_{\mbox{ }b}+\omega^{a}_{\mbox{ }c}\wedge \omega^{c}_{\mbox{ }b}
-(e^{h-g}h')^2e^a\wedge e_b\ .
\end{align}
In order to convert them into the Riemann tensor, we first note that
\be
R^{A}_{\mbox{ }B}=\half R^{A}_{\mbox{ }BCD}e^C\wedge e^D=\half R^{A}_{\mbox{ }B\mu\nu}dx^{\mu}\wedge dx^{\nu}
\qquad\mbox{and}\qquad
R^{\alpha}_{\mbox{ }\beta\mu\nu}=E_A^{\mbox{ }\alpha}e^B_{\mbox{  }\beta}R^{A}_{\mbox{ }B\mu\nu}\ .
\ee
Then the Ricci tensor is found to be
\begin{align}
R_{\beta\nu}&\equiv R^{\alpha}_{\mbox{ }\beta\alpha\nu}=E_A^{\mbox{ }\alpha}e^B_{\mbox{  }\beta}R^{A}_{\mbox{ }B\alpha\nu}
=e^{h-g}E_r^{\mbox{ }\alpha}e^b_{\mbox{  }\beta}R^{r}_{\mbox{ }b\alpha\nu}+e^{g-h}E_a^{\mbox{ }\alpha}e^r_{\mbox{  }\beta}R^{a}_{\mbox{ }r\alpha\nu}
+E_a^{\mbox{ }\alpha}e^b_{\mbox{  }\beta}R^{a}_{\mbox{ }b\alpha\nu}\nn\\
&=e^{h-g}e^b_{\mbox{  }\beta}R^{r}_{\mbox{ }br\nu}+e^{g-h}E_a^{\mbox{ }\alpha}R^{a}_{\mbox{ }r\alpha\nu}\delta_{\beta r}
+(R_3)_{\beta\nu}-2(e^{h-g}h')^2(g_3)_{\beta \nu}\ .
\end{align}
This reads
\begin{align}
R_{rr}=-3e^{g-h}\left(e^{h-g}h'\right)'\ ,\quad
R_{ij}=-e^{h-g}\left(e^{h-g}h'\right)'(g_3)_{ij}+(R_3)_{ij}-2(e^{h-g}h')^2(g_3)_{ij}\ .
\end{align}
We thus find that
\be
R=-6e^{-(g+h)}\left(e^{h-g}h'\right)'+e^{-2h}\left(R_3-6(e^{h-g}h')^2\right)\ .
\ee
As a check, in the case of AdS$_4$, since $e^{2h}=e^{-2g}=r^2-1$, we find that
\be
R_{rr}=-{3\over r^2-1}=-3g_{rr}\ ,\qquad\qquad R_{ij}=-3(r^2-1)(g_3)_{ij}=-3g_{ij}\ .
\ee
In our application the metric is parametrized by
\be
e^{2g(r)}=L^2/q(r)\ ,\qquad\qquad e^{2h(r)}=q(r)\ .
\ee
The Lagrangian for the Einstein-Hilbert action can then be calculated as
\begin{align}\label{app:Lagrangian}
{\cal L}_{\rm EH}=\sqrt{g}R&=3\cos\theta\sin\theta\left[-{qq''\over L}+{L\over 3}R_3-{q'^2\over 2L}\right]\nn\\
&=6\cos\theta\sin\theta\left[-{d\over dr}\left({qq'\over 2L}\right)+{q'^2\over 4L}+L+{L\over 3\sin\theta\cos\theta}\left({1\over n}-1\right)\delta(\theta)\right]\ .
\end{align}



\begin{thebibliography}{40}

\bibitem{Almheiri:2014lwa} 
  A.~Almheiri, X.~Dong and D.~Harlow,
  ``Bulk Locality and Quantum Error Correction in AdS/CFT,''
  JHEP {\bf 1504}, 163 (2015)
  doi:10.1007/JHEP04(2015)163
  [arXiv:1411.7041 [hep-th]].

\bibitem{Pastawski:2015qua} 
  F.~Pastawski, B.~Yoshida, D.~Harlow and J.~Preskill,
  ``Holographic quantum error-correcting codes: Toy models for the bulk/boundary correspondence,''
  JHEP {\bf 1506}, 149 (2015)
  doi:10.1007/JHEP06(2015)149
  [arXiv:1503.06237 [hep-th]].

\bibitem{Maldacena:2001kr} 
  J.~M.~Maldacena,
  ``Eternal black holes in anti-de Sitter,''
  JHEP {\bf 0304}, 021 (2003)
  doi:10.1088/1126-6708/2003/04/021
  [hep-th/0106112].

\bibitem{Ryu:2006bv} 
  S.~Ryu and T.~Takayanagi,
  ``Holographic derivation of entanglement entropy from AdS/CFT,''
  Phys.\ Rev.\ Lett.\  {\bf 96}, 181602 (2006)
  doi:10.1103/PhysRevLett.96.181602
  [hep-th/0603001].
  
\bibitem{Maldacena:1997re} 
  J.~M.~Maldacena,
  ``The Large $N$ limit of superconformal field theories and supergravity,''
  Int.\ J.\ Theor.\ Phys.\  {\bf 38}, 1113 (1999)
  [Adv.\ Theor.\ Math.\ Phys.\  {\bf 2}, 231 (1998)]
  doi:10.1023/A:1026654312961, 10.4310/ATMP.1998.v2.n2.a1
  [hep-th/9711200].
    
\bibitem{tHooft:1993dmi} 
  G.~'t Hooft,
  ``Dimensional reduction in quantum gravity,''
  Conf.\ Proc.\ C {\bf 930308}, 284 (1993)
  [gr-qc/9310026].
  
\bibitem{Susskind:1994vu} 
  L.~Susskind,
  ``The World as a hologram,''
  J.\ Math.\ Phys.\  {\bf 36}, 6377 (1995)
  doi:10.1063/1.531249
  [hep-th/9409089].
  
 
  
\bibitem{Hubeny:2007xt} 
  V.~E.~Hubeny, M.~Rangamani and T.~Takayanagi,
  ``A Covariant holographic entanglement entropy proposal,''
  JHEP {\bf 0707}, 062 (2007)
  doi:10.1088/1126-6708/2007/07/062
  [arXiv:0705.0016 [hep-th]].
  
\bibitem{Lewkowycz:2013nqa} 
  A.~Lewkowycz and J.~Maldacena,
  ``Generalized gravitational entropy,''
  JHEP {\bf 1308}, 090 (2013)
  doi:10.1007/JHEP08(2013)090
  [arXiv:1304.4926 [hep-th]].
  
\bibitem{Nishioka:2009un} 
  T.~Nishioka, S.~Ryu and T.~Takayanagi,
  ``Holographic Entanglement Entropy: An Overview,''
  J.\ Phys.\ A {\bf 42}, 504008 (2009)
  doi:10.1088/1751-8113/42/50/504008
  [arXiv:0905.0932 [hep-th]].
   
\bibitem{VanRaamsdonk:2016exw} 
  M.~Van Raamsdonk,
  ``Lectures on Gravity and Entanglement,''
  doi:10.1142/9789813149441\textunderscore 0005
  arXiv:1609.00026 [hep-th].
  
   
\bibitem{Faulkner:2013ana} 
  T.~Faulkner, A.~Lewkowycz and J.~Maldacena,
  ``Quantum corrections to holographic entanglement entropy,''
  JHEP {\bf 1311}, 074 (2013)
  doi:10.1007/JHEP11(2013)074
  [arXiv:1307.2892 [hep-th]].
  
\bibitem{Swingle:2014uza} 
  B.~Swingle and M.~Van Raamsdonk,
  ``Universality of Gravity from Entanglement,''
  arXiv:1405.2933 [hep-th].
  
\bibitem{Barrella:2013wja} 
  T.~Barrella, X.~Dong, S.~A.~Hartnoll and V.~L.~Martin,
  ``Holographic entanglement beyond classical gravity,''
  JHEP {\bf 1309}, 109 (2013)
  doi:10.1007/JHEP09(2013)109
  [arXiv:1306.4682 [hep-th]].
  
\bibitem{Halliwell:1988wc} 
  J.~J.~Halliwell,
  ``Derivation of the Wheeler-De Witt Equation from a Path Integral for Minisuperspace Models,''
  Phys.\ Rev.\ D {\bf 38}, 2468 (1988).
  doi:10.1103/PhysRevD.38.2468.
  
\bibitem{Halliwell:1988ik} 
  J.~J.~Halliwell and J.~Louko,
  ``Steepest Descent Contours in the Path Integral Approach to Quantum Cosmology. 1. The De Sitter Minisuperspace Model,''
  Phys.\ Rev.\ D {\bf 39}, 2206 (1989).
  doi:10.1103/PhysRevD.39.2206.

\bibitem{Caputa:2018asc} 
  P.~Caputa and S.~Hirano,
  ``Airy Function and 4d Quantum Gravity,''
  JHEP {\bf 1806}, 106 (2018)
  doi:10.1007/JHEP06(2018)106
  [arXiv:1804.00942 [hep-th]].
  
   
\bibitem{Caputa:2019pam} 
  P.~Caputa, S.~Datta and V.~Shyam,
  ``Sphere partition functions \& cut-off AdS,''
  JHEP {\bf 1905}, 112 (2019)
  doi:10.1007/JHEP05(2019)112
  [arXiv:1902.10893 [hep-th]].
  
\bibitem{Donnelly:2019pie} 
  W.~Donnelly, E.~LePage, Y.~Y.~Li, A.~Pereira and V.~Shyam,
  ``Quantum corrections to finite radius holography and holographic entanglement entropy,''
  arXiv:1909.11402 [hep-th].
   
   
\bibitem{Aharony:2008ug} 
  O.~Aharony, O.~Bergman, D.~L.~Jafferis and J.~Maldacena,
  ``${\cal N}=6$ superconformal Chern-Simons-matter theories, M2-branes and their gravity duals,''
  JHEP {\bf 0810}, 091 (2008)
  doi:10.1088/1126-6708/2008/10/091
  [arXiv:0806.1218 [hep-th]].
  
\bibitem{Aharony:2008gk} 
  O.~Aharony, O.~Bergman and D.~L.~Jafferis,
  ``Fractional M2-branes,''
  JHEP {\bf 0811}, 043 (2008)
  doi:10.1088/1126-6708/2008/11/043
  [arXiv:0807.4924 [hep-th]].

 

\bibitem{Fuji:2011km} 
  H.~Fuji, S.~Hirano and S.~Moriyama,
  ``Summing Up All Genus Free Energy of ABJM Matrix Model,''
  JHEP {\bf 1108}, 001 (2011)
  doi:10.1007/JHEP08(2011)001
  [arXiv:1106.4631 [hep-th]].
  
\bibitem{Marino:2011eh} 
  M.~Mari\~no and P.~Putrov,
  ``ABJM theory as a Fermi gas,''
  J.\ Stat.\ Mech.\  {\bf 1203}, P03001 (2012)
  doi:10.1088/1742-5468/2012/03/P03001
  [arXiv:1110.4066 [hep-th]].

\bibitem{Arnowitt:1959ah} 
  R.~L.~Arnowitt, S.~Deser and C.~W.~Misner,
  ``Dynamical Structure and Definition of Energy in General Relativity,''
  Phys.\ Rev.\  {\bf 116}, 1322 (1959).
  doi:10.1103/PhysRev.116.1322

\bibitem{Gibbons:1976ue} 
  G.~W.~Gibbons and S.~W.~Hawking,
  ``Action Integrals and Partition Functions in Quantum Gravity,''
  Phys.\ Rev.\ D {\bf 15}, 2752 (1977).
  doi:10.1103/PhysRevD.15.2752.
  
\bibitem{York:1972sj} 
  J.~W.~York, Jr.,
  ``Role of conformal three geometry in the dynamics of gravitation,''
  Phys.\ Rev.\ Lett.\  {\bf 28}, 1082 (1972).
  doi:10.1103/PhysRevLett.28.1082.
  
\bibitem{Hartle:1983ai} 
  J.~B.~Hartle and S.~W.~Hawking,
  ``Wave Function of the Universe,''
  Phys.\ Rev.\ D {\bf 28}, 2960 (1983).
  doi:10.1103/PhysRevD.28.2960.
  
  

\bibitem{Balasubramanian:1999re}
  V.~Balasubramanian and P.~Kraus,
  ``A Stress tensor for Anti-de Sitter gravity,''
  Commun.\ Math.\ Phys.\  {\bf 208} (1999) 413
  [hep-th/9902121].

\bibitem{Emparan:1999pm} 
  R.~Emparan, C.~V.~Johnson and R.~C.~Myers,
  ``Surface terms as counterterms in the AdS/CFT correspondence,''
  Phys.\ Rev.\ D {\bf 60}, 104001 (1999)
  doi:10.1103/PhysRevD.60.104001
  [hep-th/9903238].
  
\bibitem{deHaro:2000vlm} 
  S.~de Haro, S.~N.~Solodukhin and K.~Skenderis,
  ``Holographic reconstruction of space-time and renormalization in the AdS/CFT correspondence,''
  Commun.\ Math.\ Phys.\  {\bf 217}, 595 (2001)
  doi:10.1007/s002200100381
  [hep-th/0002230].
  

\bibitem{Drukker:2010nc} 
  N.~Drukker, M.~Mari\~no and P.~Putrov,
  ``From weak to strong coupling in ABJM theory,''
  Commun.\ Math.\ Phys.\  {\bf 306}, 511 (2011)
  doi:10.1007/s00220-011-1253-6
  [arXiv:1007.3837 [hep-th]].

\bibitem{Marino:2011nm} 
  M.~Mari\~no,
  ``Lectures on localization and matrix models in supersymmetric Chern-Simons-matter theories,''
  J.\ Phys.\ A {\bf 44}, 463001 (2011)
  doi:10.1088/1751-8113/44/46/463001
  [arXiv:1104.0783 [hep-th]].

\bibitem{Honda:2012bx} 
  M.~Honda, M.~Hanada, Y.~Honma, J.~Nishimura, S.~Shiba and Y.~Yoshida,
  ``Monte Carlo studies of 3d ${\cal N}=6$ SCFT via localization method,''
  PoS LATTICE {\bf 2012}, 233 (2012)
  doi:10.22323/1.164.0233
  [arXiv:1211.6844 [hep-lat]].
  
\bibitem{Putrov:2012zi} 
  P.~Putrov and M.~Yamazaki,
  ``Exact ABJM Partition Function from TBA,''
  Mod.\ Phys.\ Lett.\ A {\bf 27}, 1250200 (2012)
  doi:10.1142/S0217732312502008
  [arXiv:1207.5066 [hep-th]].

 
\bibitem{Drukker:2011zy} 
  N.~Drukker, M.~Mari\~no and P.~Putrov,
  ``Nonperturbative aspects of ABJM theory,''
  JHEP {\bf 1111}, 141 (2011)
  doi:10.1007/JHEP11(2011)141
  [arXiv:1103.4844 [hep-th]].
  
\bibitem{Hatsuda:2012dt} 
  Y.~Hatsuda, S.~Moriyama and K.~Okuyama,
  ``Instanton Effects in ABJM Theory from Fermi Gas Approach,''
  JHEP {\bf 1301}, 158 (2013)
  doi:10.1007/JHEP01(2013)158
  [arXiv:1211.1251 [hep-th]];
  ``Instanton Bound States in ABJM Theory,''
  JHEP {\bf 1305}, 054 (2013)
  doi:10.1007/JHEP05(2013)054
  [arXiv:1301.5184 [hep-th]].
  
\bibitem{Hatsuda:2013oxa} 
  Y.~Hatsuda, M.~Mari\~no, S.~Moriyama and K.~Okuyama,
  ``Non-perturbative effects and the refined topological string,''
  JHEP {\bf 1409}, 168 (2014)
  doi:10.1007/JHEP09(2014)168
  [arXiv:1306.1734 [hep-th]].

\bibitem{Codesido:2014oua} 
  S.~Codesido, A.~Grassi and M.~Mari\~no,
  ``Exact results in ${\cal N}=8$ Chern-Simons-matter theories and quantum geometry,''
  JHEP {\bf 1507}, 011 (2015)
  doi:10.1007/JHEP07(2015)011
  [arXiv:1409.1799 [hep-th]].
    
\bibitem{Vilenkin:1986cy} 
  A.~Vilenkin,
  ``Boundary Conditions in Quantum Cosmology,''
  Phys.\ Rev.\ D {\bf 33}, 3560 (1986).
  doi:10.1103/PhysRevD.33.3560
    
\bibitem{Bergman:2009zh}
  O.~Bergman, S.~Hirano,
  ``Anomalous radius shift in AdS$_4$/CFT$_3$,''
  JHEP {\bf 0907}, 016 (2009).
  [arXiv:0902.1743 [hep-th]].

\bibitem{Aharony:2009fc}
  O.~Aharony, A.~Hashimoto, S.~Hirano and P.~Ouyang,
  ``D-brane Charges in Gravitational Duals of 2+1 Dimensional Gauge Theories
  and Duality Cascades,''
  JHEP {\bf 1001}, 072 (2010)
  [arXiv:0906.2390 [hep-th]].

\bibitem{Dabholkar:2014wpa} 
  A.~Dabholkar, N.~Drukker and J.~Gomes,
  ``Localization in supergravity and quantum AdS$_4$/CFT$_3$ holography,''
  JHEP {\bf 1410}, 90 (2014)
  doi:10.1007/JHEP10(2014)090
  [arXiv:1406.0505 [hep-th]].
 


\bibitem{deBoer:1999tgo} 
  J.~de Boer, E.~P.~Verlinde and H.~L.~Verlinde,
  ``On the holographic renormalization group,''
  JHEP {\bf 0008}, 003 (2000)
  doi:10.1088/1126-6708/2000/08/003
  [hep-th/9912012].
  
\bibitem{McGough:2016lol} 
  L.~McGough, M.~Mezei and H.~Verlinde,
  ``Moving the CFT into the bulk with $T\bar T$,''
  arXiv:1611.03470 [hep-th].

\bibitem{Solodukhin:1994yz} 
  S.~N.~Solodukhin,
  ``The Conical singularity and quantum corrections to entropy of black hole,''
  Phys.\ Rev.\ D {\bf 51}, 609 (1995)
  doi:10.1103/PhysRevD.51.609
  [hep-th/9407001].
  
\bibitem{Nishioka:2013haa} 
  T.~Nishioka and I.~Yaakov,
  ``Supersymmetric Renyi Entropy,''
  JHEP {\bf 1310}, 155 (2013)
  doi:10.1007/JHEP10(2013)155
  [arXiv:1306.2958 [hep-th]].
  
\bibitem{Nishioka:2014mwa} 
  T.~Nishioka,
  ``The Gravity Dual of Supersymmetric Renyi Entropy,''
  JHEP {\bf 1407}, 061 (2014)
  doi:10.1007/JHEP07(2014)061
  [arXiv:1401.6764 [hep-th]].
  
\bibitem{Casini:2010kt} 
  H.~Casini and M.~Huerta,
  ``Entanglement entropy for the $n$-sphere,''
  Phys.\ Lett.\ B {\bf 694}, 167 (2011)
  doi:10.1016/j.physletb.2010.09.054
  [arXiv:1007.1813 [hep-th]];
  ``Entanglement entropy in free quantum field theory,''
  J.\ Phys.\ A {\bf 42}, 504007 (2009)
  doi:10.1088/1751-8113/42/50/504007
  [arXiv:0905.2562 [hep-th]]; 
  H.~Casini, M.~Huerta and R.~C.~Myers,
  ``Towards a derivation of holographic entanglement entropy,''
  JHEP {\bf 1105}, 036 (2011)
  doi:10.1007/JHEP05(2011)036
  [arXiv:1102.0440 [hep-th]].
  
 
\bibitem{Maldacena:1998im} 
  J.~M.~Maldacena,
  ``Wilson loops in large $N$ field theories,''
  Phys.\ Rev.\ Lett.\  {\bf 80}, 4859 (1998)
  doi:10.1103/PhysRevLett.80.4859
  [hep-th/9803002].
  
\bibitem{Rey:1998ik} 
  S.~J.~Rey and J.~T.~Yee,
  ``Macroscopic strings as heavy quarks in large $N$ gauge theory and anti-de Sitter supergravity,''
  Eur.\ Phys.\ J.\ C {\bf 22}, 379 (2001)
  doi:10.1007/s100520100799
  [hep-th/9803001].
    
\bibitem{Drukker:2008zx} 
  N.~Drukker, J.~Plefka and D.~Young,
  ``Wilson loops in 3-dimensional ${\cal N}=6$ supersymmetric Chern-Simons Theory and their string theory duals,''
  JHEP {\bf 0811}, 019 (2008)
  doi:10.1088/1126-6708/2008/11/019
  [arXiv:0809.2787 [hep-th]].
  
\bibitem{Chen:2008bp} 
  B.~Chen and J.~B.~Wu,
  ``Supersymmetric Wilson Loops in ${\cal N}=6$ Super Chern-Simons-matter theory,''
  Nucl.\ Phys.\ B {\bf 825}, 38 (2010)
  doi:10.1016/j.nuclphysb.2009.09.015
  [arXiv:0809.2863 [hep-th]].
  
\bibitem{Rey:2008bh} 
  S.~J.~Rey, T.~Suyama and S.~Yamaguchi,
  ``Wilson Loops in Superconformal Chern-Simons Theory and Fundamental Strings in Anti-de Sitter Supergravity Dual,''
  JHEP {\bf 0903}, 127 (2009)
  doi:10.1088/1126-6708/2009/03/127
  [arXiv:0809.3786 [hep-th]].
  
\bibitem{Klemm:2012ii} 
  A.~Klemm, M.~Mari\~no, M.~Schiereck and M.~Soroush,
  ``Aharony-Bergman-Jafferis-Maldacena Wilson loops in the Fermi gas approach,''
  Z.\ Naturforsch.\ A {\bf 68}, 178 (2013)
  doi:10.5560/ZNA.2012-0118
  [arXiv:1207.0611 [hep-th]].
 
  
\bibitem{Fujita:2009kw} 
  M.~Fujita, W.~Li, S.~Ryu and T.~Takayanagi,
  ``Fractional Quantum Hall Effect via Holography: Chern-Simons, Edge States, and Hierarchy,''
  JHEP {\bf 0906}, 066 (2009)
  doi:10.1088/1126-6708/2009/06/066
  [arXiv:0901.0924 [hep-th]].
  
\bibitem{Bhattacharyya:2012ye} 
  S.~Bhattacharyya, A.~Grassi, M.~Marino and A.~Sen,
  ``A One-Loop Test of Quantum Supergravity,''
  Class.\ Quant.\ Grav.\  {\bf 31}, 015012 (2014)
  doi:10.1088/0264-9381/31/1/015012
  [arXiv:1210.6057 [hep-th]].
  
  
\bibitem{Engelhardt:2014gca} 
  N.~Engelhardt and A.~C.~Wall,
  ``Quantum Extremal Surfaces: Holographic Entanglement Entropy beyond the Classical Regime,''
  JHEP {\bf 1501}, 073 (2015)
  doi:10.1007/JHEP01(2015)073
  [arXiv:1408.3203 [hep-th]].
  
\bibitem{Benini:2015eyy} 
  F.~Benini, K.~Hristov and A.~Zaffaroni,
  ``Black hole microstates in AdS$_{4}$ from supersymmetric localization,''
  JHEP {\bf 1605}, 054 (2016)
  doi:10.1007/JHEP05(2016)054
  [arXiv:1511.04085 [hep-th]].

\bibitem{Zaffaroni:2019dhb} 
  A.~Zaffaroni,
  ``Lectures on AdS Black Holes, Holography and Localization,''
  arXiv:1902.07176 [hep-th].

\end{thebibliography}
\end{document}